\documentclass[aip,amsmath,amssymb,reprint,]{revtex4-1}

\usepackage{amsmath}
\usepackage{graphicx}
\usepackage{bm}
\usepackage{xr}

\begin{document}
\title{Atomistic Generative Diffusion for Materials Modeling}

\author{Nikolaj Rønne}
\email{niron@dtu.dk}
\affiliation{Department of Energy Conversion and Storage, Technical University of Denmark, DK 2800 Kgs. Lyngby, Denmark}
\affiliation{CAPeX Pioneer Center for Accelerating P2X Materials Discovery, DK 2800 Kgs. Lyngby, Denmark}
\author{Bjørk Hammer}
\affiliation{Center for Interstellar Catalysis, Department of Physics and Astronomy, Aarhus University, DK‐8000 Aarhus C, Denmark}

\date{\today}

\begin{abstract}
We present a generative modeling framework for atomistic systems that
combines score-based diffusion for atomic positions with a novel
continuous-time discrete diffusion process for atomic types. This
approach enables flexible and physically grounded generation of atomic
structures across chemical and structural domains. Applied to metallic
clusters and two-dimensional materials using the QCD and C2DB datasets, our models
achieve strong performance in fidelity and diversity, evaluated using
precision–recall metrics against synthetic baselines. We demonstrate
atomic type interpolation for generating bimetallic clusters beyond
the training distribution, and use classifier-free guidance to steer
sampling toward specific crystallographic symmetries in two-dimensional
materials. These capabilities are implemented in Atomistic Generative
Diffusion (AGeDi), an open-source, extensible software package for atomistic generative
diffusion modeling. 
\end{abstract}

\maketitle

\section{Introduction}
\label{sec:introduction}
Machine learning methods have become central to modern materials
science, particularly through the development of machine-learned
interatomic potentials\cite{behler2007,bartok2010} (MLIPs) that significantly improve the accuracy,
speed, and scalability of atomistic simulations. By learning the
potential energy landscapes from quantum mechanical data, MLIPs have made it
possible to access larger system sizes, longer time scales, and
greater chemical diversity than what is feasible with traditional ab
initio methods. These models have enabled advances across domains
including molecular dynamics, structure prediction, catalysis, and
materials design.\cite{deringer2019,friederich2021}

The rapid progress in MLIPs has been driven by both algorithmic
innovations — such as message-passing neural networks\cite{schutt2018,unke2019,gasteiger2019,chen2019}, equivariant
models\cite{anderson2019,gasteiger2021,schutt2021,batzner2022,batatia2022,haghighatlari2022}, and active learning\cite{vandermause2020,jinnouchi2020,novikov2020,ronne2022,kulichenko2023} — and the emergence of accessible software
ecosystems that lower the barrier to entry for the broader materials
science community.

Generative diffusion models have recently opened
new frontiers in 
atomistic modeling\cite{park2024}, with promising applications including crystal
structure prediction\cite{xie2022,lyngby2022,jiao2023,zhao2023,yang2024,luo2024,alverson2024,kwon2024,levy2024,zhong2025,joshi2025}, efficient sampling of complex energy
landscapes\cite{ronne2024,park2024a}, and multi-objective materials
optimization\cite{zeni2025,jia2025,yang2025,park2025,guo2025}. The
ability to generate realistic atomic configurations with target 
properties has far-reaching implications for materials
discovery. Generative models can accelerate the identification of
stable crystal structures, support the inverse design of functional
materials, and enable efficient exploration of complex potential
energy surfaces that are difficult to sample using classical
methods. Applications such as high-entropy alloy design, heterogeneous
catalysis, and energy storage materials discovery benefit particularly from
models that generate diverse and physically valid configurations. By
enabling conditional generation, diffusion models offer a path toward
data-driven design workflows, replacing traditional trial-and-error
methods with guided, generative exploration.\cite{kwon2024}

In this work, we develop a unified generative modeling framework for
atomistic systems that combines score-based diffusion for atomic
positions with a novel continuous-time discrete diffusion process for
atomic types. This approach enables simultaneous and physically
consistent generation of structure and composition. We evaluate the
framework on two representative datasets: the Quantum Cluster Database\cite{manna2023}
(QCD) for metallic nanoclusters and the Computational 2D Materials
Database\cite{haastrup2018} (C2DB) for two-dimensional crystals. In both cases, the
generative models exhibit strong fidelity and diversity, assessed
using a precision–recall–based evaluation metric with synthetic
baselines. Beyond unconditional generation, we demonstrate possible
capabilities of the approach: interpolation over atomic types to
sample bimetallic clusters outside the training distribution, and
classifier-free guidance to generate two-dimensional materials with targeted
crystallographic symmetries. These capabilities are implemented in
AGeDi, an open-source, extensible software package designed to support
generative diffusion modeling in atomistic materials discovery
workflows. 

This paper begins in Section~\ref{sec:methods} with the theoretical
foundations of our generative framework, covering diffusion over
atomic positions and types, conditional generation via classifier-free
guidance, and type interpolation. Section~\ref{sec:agedi} introduces
the AGeDi software architecture. Section~\ref{sec:eval} describes our
evaluation approach using precision–recall metrics and synthetic
baselines. In Section~\ref{sec:results}, we demonstrate AGeDi on
metallic clusters and 2D materials, highlighting compositional
interpolation and symmetry-guided
sampling. Section~\ref{sec:discussion} outlines broader implications
and future directions, and Section~\ref{sec:conclusion} concludes.

\section{Methods}
\label{sec:methods}
Generative diffusion models are a class of deep learning approaches
that have gained significant attention for their ability to generate
high-quality and diverse samples. They operate based on principles
from non-equilibrium thermodynamics, in which data is gradually noised
during a forward process, and then recovered by reversing this process
during sampling. A diffusion model consists of a forward and a reverse process. The
forward diffusion process adds noise to the data until it
approximates a tractable prior distribution. This forward process is
analytically defined. In contrast, the reverse process is a learned
denoising procedure that gradually removes noise to recover data from
the underlying distribution. 

Training a diffusion model involves defining the forward and reverse
process and optimizing a score network, which predicts the noise
added during forward diffusion. This is done by applying the forward
noise process to training data, using the score network to predict the
added noise, and updating the network via a loss function that
penalizes the discrepancy between the predicted and true noise. 

Sampling constitutes the generative phase, where new data is produced
by denoising samples drawn from the prior distribution. The reverse
diffusion process transforms these noisy samples into realistic
configurations by approximating the true data distribution. Rather
than memorizing individual examples, the model learns the underlying
structure and statistical patterns of the data.

An atomistic system with $N$ atoms is represented by the tuple:
\begin{equation}
  \mathcal{M} = (\mathbf{R}, \mathbf{z}, \mathbf{S}),
\end{equation}
where $\mathbf{R} \in \mathbb{R}^{3\times N}$ represents atomic
positions, $\mathbf{z} \in \mathbb{Z}^{N}$ are the atomic types and
$\mathbf{S} \in \mathbb{R}^{3 \times 3}$ is the periodic cell.
Each of these variables can undergo diffusion, but they impose different
modeling challenges due to their distinct properties.

Atomic positions, $\mathbf{R}$, are equivariant under rotation and translation, and
are coupled with the periodic cell, $\mathbf{S}$, which defines boundary
conditions. In contrast, atomic types, $\mathbf{z}$, are discrete variables, and
require a distinct modeling approach within the diffusion framework.

In the following, we present the theoretical foundations
of score-based diffusion and continuous-time discrete diffusion as
applied to atomistic systems. In this work, we limit ourselves to
diffusion of atomic positions and types.

\subsection{Score-based Diffusion through Stochastic Differential
  Equations}
Following Song et al.\cite{song2021}, we define a generative diffusion model as a stochastic process that
gradually transforms samples from the data distribution into a
tractable prior through a forward diffusion process, and reverses this
transformation via a learned denoising process.

Let $\mathcal{M}_0 \sim p_0(\mathcal{M})$ denote a sample from the data
distribution - that is, an atomic configuration - and $\mathbf{R}_0$
the corresponding atomic positions. In the following, we present the
theory for diffusion of atomic positions and in section
\ref{ssec:z-diffusion} we present discrete atomic type diffusion.
We define the forward process as a stochastic process
$\{\mathbf{R}_t\}_{t=0}^{1}$ indexed by a continuous time
variable $t\in [0, 1]$ transforming a sample from the data
distribution $\mathbf{R}_0$ into a sample from a tractable prior
$\mathbf{R}_1 \sim p_1$ given by the stochastic differential
equation (SDE)
\begin{equation}\label{eq:forward-sde}
  d\mathbf{R}_t = \mathbf{f}(\mathbf{R}_t, t)dt + g(t)d\mathbf{W}_t,
\end{equation}
where $\mathbf{f}(\cdot, t)$ represents the drift term, $g(t)$ the
diffusion coefficient, and $\mathbf{W}_t$ denotes Wiener noise, all to
be detailed below.

The reverse time process governing generation is given by the SDE
\begin{equation}\label{eq:reverse-sde}
  d\mathbf{R}_t = [\mathbf{f}(\mathbf{R}_t, t) - g^2(t)
  \nabla_{\mathbf{R}_t} \log p_t(\mathbf{R}_t) ]dt + g(t)d\mathbf{\overline{W}}_t,
\end{equation}
where $\nabla_{\mathbf{R}_t} \log p_t(\mathbf{R}_t)$ is the score
function and $\mathbf{\overline{W}}_t$ is the reverse time Wiener noise. 

Since $\nabla_{\mathbf{R}_t} \log p_t(\mathbf{R}_t)$ is not known, we
wish to learn a time-conditioned score network $\mathbf{s}_\theta(\mathcal{M}_t, t)$ that
estimates the true score function using data from the data
distribution. The score network is typically implemented as a graph
neural network (GNN) taking an atomic structure, $\mathcal{M}$, as input and
predicting the score. This is achieved through denoising score-matching with
the optimization objective given as
\begin{eqnarray}
  \underset{\theta}{\mathrm{argmin}} \; \mathbb{E}_t\bigg\{&&\lambda(t)
  \mathbb{E}_{\mathbf{R}_0}\mathbb{E}_{\mathbf{R}_t|\mathbf{R}_0}
  \Big[||\mathbf{s}_\theta(\mathcal{M}_t, t) -\nonumber\\ 
   && \nabla_{\mathbf{R}_t} \log p_{0t}(\mathbf{R}_t| \mathbf{R}_0) ||_2^2\Big] \bigg\}.
\end{eqnarray}
$\lambda(t)$ denotes a positive weighting function, $t$ is uniformly
sampled over $[0,1]$, $\mathbf{R}_0 \sim p_0(\mathbf{R})$ and
$\mathbf{R}_t \sim p_{0t}(\mathbf{R}_t| \mathbf{R}_0)$, where
$p_{0t}(\cdot| \cdot)$ denotes the transition kernel for
noising data.

Modeling atomistic systems, we need to ensure the score network obeys
the symmetries of the system. Specifically for position diffusion,
$\nabla_{\mathbf{R}_t} \log p_{0t}(\mathbf{R}_t| \mathbf{R}_0)$
represents a force field with the same dimensionality as the positions variable ($\mathbb{R}^{N\times 3}$). We can exploit the
developments in equivariant MLIPs and use an equivariant GNN with multiple prediction heads e.g. an 
equivariant head that predicts a non-conservative force, to model the
positions score function.

In order to proceed further, the SDE must be detailed explicitly,
i.e. choices for $\mathbf{f}$, $\mathbf{g}$ and $\mathbf{W}$ must be
made. In AGeDi we adopt two widely used SDE formulations — the Variance Preserving (VP) and
Variance Exploding (VE) processes. The VP SDE is given as
\begin{equation}\label{eq:vp-sde}
  d\mathbf{R}_t = -\frac{1}{2} \beta(t) \mathbf{R}_t dt + \sqrt{\beta(t)}d\mathbf{W}_t,
\end{equation}
where $\beta(t)$ is an increasing function denoting the noise
scheduling. The VP transition kernel
is Gaussian and given as
\begin{equation}
  p_{0t}(\mathbf{R}_t| \mathbf{R}_0) = \mathcal{N}(\mathbf{R}_t;
  \mathbf{R}_0 e^{-\frac{1}{2}\alpha(t)}, \mathbb{I}(1-e^{-\alpha(t)})),
\end{equation}
where $\alpha(t) = \int_0^t \beta(s)ds$. This simplifies the denoising
training objective significantly since
\begin{equation}\label{eq:grad-log-transition}
  \nabla_{\mathbf{R}_t} \log p_{0t}(\mathbf{R}_t| \mathbf{R}_0) = \frac{-\mathbf{W}}{\varsigma_t}
\end{equation}
for $\mathbf{W} \sim \mathcal{N}(0,1)$ and $\varsigma_t^2 = 1-e^{-\alpha(t)}$.\cite{song2021} Hereby, it is clear that the score network is
trained to directly predict the noise added to the
training data. 

One obtains a similar result for the VE SDE, that is given through
\begin{equation}\label{eq:vp-sde}
  d\mathbf{R}_t = \sqrt{\frac{d[\sigma^2(t)]}{dt}}d\mathbf{W}_t,
\end{equation}
where $\sigma(t)$ again denotes the noise schedule. Notice VE does
have zero drift. For the VE SDE
transition kernel one gets the same result for the denoising score
matching objective as Eq.~\ref{eq:grad-log-transition} but with
$\varsigma_t^2 = \sigma^2(t) - \sigma^2(0)$. 

Once the score network $\mathbf{s}_\theta(\mathcal{M}_t, t)$ has been
trained to approximate the score function,
$\nabla_{\mathbf{R}_t} \log p_t(\mathbf{R}_t)$, it can be used to generate new atomistic
configurations by simulating the reverse SDE given by
Eq.~\ref{eq:reverse-sde}. For practical sampling, this SDE is
discretized using numerical solvers. One of the most widely used methods is the
Euler-Maruyama scheme. Here, we approximate Eq.~\ref{eq:reverse-sde}
by a discrete update rule given by
\begin{eqnarray}
  \mathbf{R}_{t_{i-1}} =&& \mathbf{R}_{t_i} +\nonumber \\
  &&\left[\mathbf{f}(\mathbf{R}_{t_i}, t_i) - g(t_i)^2 \,
    \mathbf{s}_\theta(\mathcal{M}_{t_i}, t_i)\right] \Delta t
  +\nonumber \\
  &&g(t_i) \sqrt{|\Delta t|} \cdot \mathbf{W}_i,
\end{eqnarray}
where $\{t_i\}_{i=0}^N$ is a decreasing sequence of time steps from
$t_N = 1$ to $t_0 = 0$, $\Delta t = t_{i-1} - t_i < 0$ is the
(negative) step size, $\mathbf{W}_i$
follows the noise distribution used in the forward process.

Sampling is performed by initializing $\mathbf{R}_1$
from the prior distribution and iteratively applying the update rule until
$t = 0$, yielding a final sample $\mathbf{R}_0$ from the learned data
distribution.

Both the VP and VE SDE formulation are implemented in AGeDi together
with multiple noise schedules. In this work, we demonstrate results
based on the VE SDE formulation with a linear noise schedule.

\subsection{Continuous-Time Discrete Diffusion for Atomic Types}
\label{ssec:z-diffusion}
In modeling atomistic systems, atomic types are inherently discrete,
taking values from a finite set of elements. To enable generative
modeling of discrete variables with a diffusion process, we
follow the framework introduced by Lou et al.~\cite{lou2024},
which introduces a novel score entropy loss that naturally extends
score matching to discrete spaces.

Let $z_0 \sim p_0(z)$ denote an 
atomic type vector over the finite support $\mathcal{Z} = \{1,2,...,z_{\mathrm{max}} \}$
from the empirical data distribution. The forward process
defines a continuous-time Markov process given by 
\begin{equation}
    \frac{dp_t}{dt} = \mathbf{Q}_t p_t,
\end{equation}
where $\mathbf{Q}_t \in \mathbb{R}^{|\mathcal{Z}|\times |\mathcal{Z}|}$ is the diffusion matrix satisfying
column-wise conservation (i.e., each column sums to zero). This
forward process gradually transitions the atomic type variables toward
a tractable prior distribution, $p_1$, for $t \rightarrow 1$.

The reverse process is analogous to the score-based diffusion process
introduced in the previous section and is given by
\begin{equation}
    \frac{dp_{t}}{dt} = \bar{\mathbf{Q}}_t p_t,
\end{equation}
where 
\begin{align}
  \bar{\mathbf{Q}}_{t,zz'} &= \frac{p_t(z)}{p_t(z')}
  \mathbf{Q}_{t,zz'} \; \; \mathrm{for}\, z \neq z' \, \mathrm{and} \\
  \bar{\mathbf{Q}}_{t,zz} &= -\sum_{z'\neq z} \mathbf{Q}_{t,z'z},
\end{align}
for any two atomic types $z$ and $z'$. The ratio $\frac{p_t(z)}{p_t(z')}$ is known as
the concrete score that generalizes the typical score function,
$\nabla_{x} \log p_t(x)$, by comparing the relative likelihoods of
atomic types to favor more probable type reconstruction during sampling.

Similar to the standard diffusion models, the
objective of discrete diffusion is to learn the concrete scores
through a score model, $\mathbf{s}_\theta$, in order to sample starting from the prior distribution.

The score-entropy optimization objective used to train the score model
is given by
\begin{eqnarray}               %
  \underset{\theta}{\mathrm{argmin}} \; \mathbb{E}_t
  \Bigg\{&&
  \mathbb{E}_{z_0}\mathbb{E}_{z_t|z_0}
  \bigg[
    \sum_{z' \neq z_t} w_{z_t z'} 
  \Big(\mathbf{s}_\theta(\mathcal{M}_t,t)_{z'} - \nonumber \\
    && \frac{p_{0t}(z' | z_0)}{p_{0t}(z_t | z_0)}\log \mathbf{s}_\theta(\mathcal{M}_t,t)_{z'}
    \Big)
    \bigg]
  \Bigg\},
\end{eqnarray}
where $w_{z_tz'}$ serves a similar purpose as
$\lambda(t)$ in score-based diffusion models, $t$ is uniformly sampled over
$[0, 1]$, $z_0 \sim p_0$ and $z_t \sim p_{0t}(z_t | z_0)$, where $p_{0t}$ denotes the
transition kernel for noising the types.

To apply noise in continuous time, we define the time-dependent diffusion matrix as
\begin{equation}
  \mathbf{Q}_t = \sigma(t) \mathbf{Q},
\end{equation}

where $\sigma(t)$ is the time-dependent noise schedule and can be chosen similar to score-based diffusion and
$\mathbf{Q}$ is a square matrix that is chosen to reflect the prior
distribution. Hereby, the transition kernel becomes
\begin{equation}
  p_{0t}(z_t | z_0) = \exp \left(\int_0^t \sigma(s) ds \mathbf{Q}\right)_{z_0},
\end{equation}
where the subscript $z_0$ indicates taking the $z_0$ column of the
matrix.

We use the absorbing state formulation\cite{lou2024}, where $\mathbf{Q}$ is given as
\begin{equation}
\mathbf{Q}_{ij} =
\begin{cases}
1, & \text{if } i = 0 \text{ and } j \neq 0 \\
-1, & \text{if } i = j \neq 0 \\
0, & \text{otherwise}
\end{cases}
\end{equation}
In this setup, we introduce a special masked atom type to characterize
atoms without a physical atom type. The masked atom type is assigned
atomic number $0$ and corresponds to the first row (and column) of the 
matrix $\mathbf{Q}$. The prior distribution at $t = 1$
is defined such that all atoms have this masked atom type. To enforce
this behavior, $\mathbf{Q}$ is constructed so that forward transitions are
only allowed from physical atom types into the
masked atom type, and not between any other types. Concretely, all
off-diagonal elements of $\mathbf{Q}$ are zero, 
except those in the first row, which allows the transitions into the masked state. This
structure ensures that once an atom has the masked atom type, it remains masked, and
no transitions occur between distinct physical atom types during the
forward diffusion process. 

In practice, during training, atomic types are progressively corrupted
by replacing them with the special masked type. At early diffusion times, most atoms
retain their original types; at later times, a larger fraction are
replaced with the masked type. Intuitively, the model is trained to predict the
original atomic type for each atom. During sampling, the process is
reversed. The model starts with all 
atomic types in masked atom type and denoises them step by step by
predicting, at each time step, the likely type of each atom.
These predictions are used to probabilistically sample unmasked
atomic types according to the learned reverse process. As sampling
progresses, more types are filled in with concrete atomic types until
a fully specified configuration is produced. Because the forward
process only replaces types with the mask token (rather than also
switching between elements), the model learns to reconstruct types
from uncertainty, not from confusion — resulting in a more stable and
interpretable generative process. 

Following Lou et al.\cite{lou2024} we use a discrete $\tau$-leaping
sampler, that performs an Euler step at each type positions
simultaneously.

\subsection{Joint Diffusion}
Once the diffusion processes for atomic positions and types have been
defined, they can be combined into a joint generative model
that produces fully specified atomistic structures. Since positions
and types evolve under different dynamics - continuous stochastic
differential equations for positions and discrete diffusion for types
- they are treated as independent variables during
both training and sampling, but noised and denoised simultaneously
during both training and sampling.

The generative model learns separate score models implemented as
independent prediction heads for each
variable, and training proceeds by minimizing the sum of the
corresponding score-matching losses: 
\begin{equation}
\mathcal{L}_{\text{total}} = \lambda_{\mathbf{R}} \mathcal{L}^{\mathbf{R}} + \lambda_{\mathbf{z}} \mathcal{L}^{\mathbf{z}},
\end{equation}
where the weights, $\lambda$, control the trade-off between learning
accurate structures and compositions. 

During sampling, both diffusion processes are denoised in parallel
using their respective learned score models. This joint formulation
ensures that positional and compositional aspects of the atomic
configuration are generated coherently, while allowing each to follow
its own diffusion schedule and dynamics. While diffusion of the
periodic cell is not included in this work, the methodology can
be expanded to also encompass diffusion of the periodic cell.

\subsection{Classifier-Free Guidance for Conditional Generation}

In generative modeling, it is often necessary to sample
configurations conditioned on target properties in order to steer the
sampling towards specific regions of the sampling space. Classifier-free
guidance\cite{ho2022,nichol2022} (CFG) provides a simple yet powerful
mechanism for enabling such conditional generation without the need
for external models.

To apply CFG, the score model is trained to
predict both conditional and unconditional score functions. This is
achieved by randomly masking the conditioning variable $\mathbf{y}$
during training with some probability $\tau$. When the condition is
present, the model learns the conditional score $\nabla_{\mathbf{R}_t}
\log p_t(\mathbf{R}_t \mid \mathbf{y})$, and when the condition is
masked (by replacing $\mathbf{y}$ with a zero
vector, $\varnothing$), it learns the unconditional score
$\nabla_{\mathbf{R}_t} \log p_t(\mathbf{R}_t)$. This enables
property-targeted sampling using only a single unified model. 

During sampling, both the conditional and unconditional scores are
evaluated and interpolated to steer the generation toward the desired
condition. This is done by modifying the score used in the
reverse-time SDE with the guided score 
\begin{equation}
  \tilde{\mathbf{s}}_\theta(\mathcal{M}_t, t, \mathbf{y}) =
    w \mathbf{s}_\theta(\mathcal{M}_t, t, \mathbf{y}) +
   (1-w)\mathbf{s}_\theta(\mathcal{M}_t, t, \varnothing),
\end{equation}
where $\mathbf{s}_\theta(\mathcal{M}_t, t, \mathbf{y})$ is the
conditional score, $\mathbf{s}_\theta(\mathcal{M}_t, t, \varnothing)$
is the unconditional score, and $w$ is the guidance scale that
determines the strength of the conditioning signal. This interpolated
score is then inserted into the numerical SDE solver, to perform
sampling under the influence of the target property. In the context of
atomistic systems, CFG offers a 
flexible and interpretable way to control generative outputs.

\subsection{Interpolating Atomic Types}
\label{sec:interpolation}
Interpolation is a common technique for generative models to sample
between training data domains allowing for exploration of unseen
samples. This will be explored for the AGeDi model trained on the QCD dataset, since only
mono-metallic clusters are part of the training data, but
interpolation will allow sampling bimetallic cluster based on the QCD
trained model.

\begin{figure}
	\centering
	\includegraphics[]{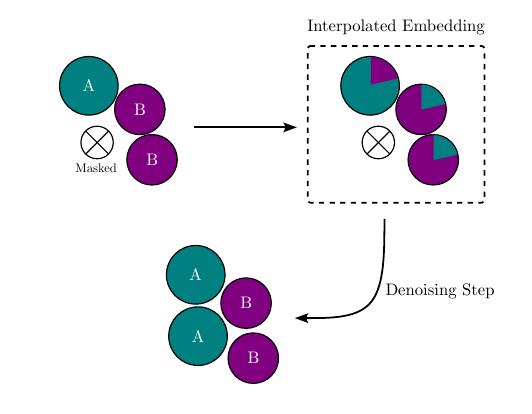}
	\caption{Conceptual illustration of the interpolation between
        two types during denoising.}
	\label{fig:interpolation-schematic}
\end{figure}

In the standard type diffusion, each atomic species is
represented by a learned embedding vector, $\mathcal{E}$. These embeddings are used
as input to the score network and encode chemically meaningful
differences between elements (e.g., Cu, Pd, Pt). During sampling,
these embeddings are treated as fixed — the model predicts 
scores based on each atom’s current type embedding. To enable
interpolation between atomic species, we modify this setup. Instead of
assigning each atom its corresponding embedding, we construct a weighted linear
combination of two atomic type embeddings. Let $\mathcal{E}_A$ and
$\mathcal{E}_B$ denote the learned embeddings for atomic types A and
B. At each diffusion step, we define an interpolated embedding:
\begin{equation}
  \tilde{\mathcal{E}}_{A/B} = \alpha \mathcal{E}_{A/B} + (1-\alpha) \mathcal{E}_{B/A},
\end{equation}
where $\alpha \sim \mathcal{U}(0,1)$ and is picked each time the interpolation
embedding is called and $\mathcal{E}_{A/B}$ refers to either embedding
A or B. See Fig.~\ref{fig:interpolation-schematic} for a conceptual
illustration of the atomic type interpolation scheme. The
interpolated embedding is applied for both types during
type score prediction when performing interpolation, while the 
diffusion process is constrained to only transition to types A and
B. The position score model remains unchanged, allowing it to
naturally extrapolate structure patterns from the training data.

The interpolation strategy effectively introduces a controlled
ambiguity into the generative process by presenting the model with
input embeddings that lie between those of two real atomic
species. This \textit{in-between} embedding does not correspond to any
specific element in the training data, but instead acts as a soft
prior — signaling to the model that either of the two interpolated
elements would be chemically plausible for a given atom. During
the reverse diffusion process, the model uses this signal to resolve
the uncertainty by probabilistically selecting one of the two real
atomic types, in a way that is structurally and chemically consistent
with the surrounding environment. This inference is guided by the
model's learned representation of local atomic configurations, which
it leverages to assign species in a physically meaningful way.

By applying this interpolation across all atoms in a
structure, the model can generate systems that exhibit mixed-element
or bimetallic character, even though such compositions were entirely
absent from the training distribution. This is made possible because
the model has learned a rich, continuous representation of local
atomic environments — one that captures the geometric and chemical
regularities present in the training data. As an example then, even though only
mono-metallic clusters are observed during training, the positional
diffusion network has internalized how atomic arrangements respond to
different chemical species, including how bond lengths, coordination
patterns, and structural motifs vary across elements. 

When presented with an interpolated type embedding during sampling,
the model interprets it not as noise, but as a chemically plausible
but uncertain atomic identity. It then relies on the learned
structure–composition correlations to infer which of the two candidate
species (from the interpolation) would best fit within the local
atomic environment — balancing geometric compatibility, bonding
patterns, and global stability. In this way, the structural score
network effectively extrapolates beyond its training domain: while it
has never seen bimetallic clusters explicitly, it can assemble them by
recombining mono-metallic motifs and adjusting them in a way that
remains physically coherent. 

Crucially, the discrete diffusion process ensures that all atomic
types in the final output correspond to valid chemical species.
The model is not inventing new elements; rather, it
is using interpolation in latent space as a controlled mechanism for
compositionally diversifying the output. This combination of
architectural flexibility, chemical priors, and representational
continuity provides a simple yet effective pathway for exploring new
regions of compositional space — enabling, for instance, the
generation of plausible bimetallic clusters from a model trained
exclusively on mono-metallic data.

\subsection{The AGeDi Software Package}
\label{sec:agedi}

\begin{figure*}
	\centering
	\includegraphics[]{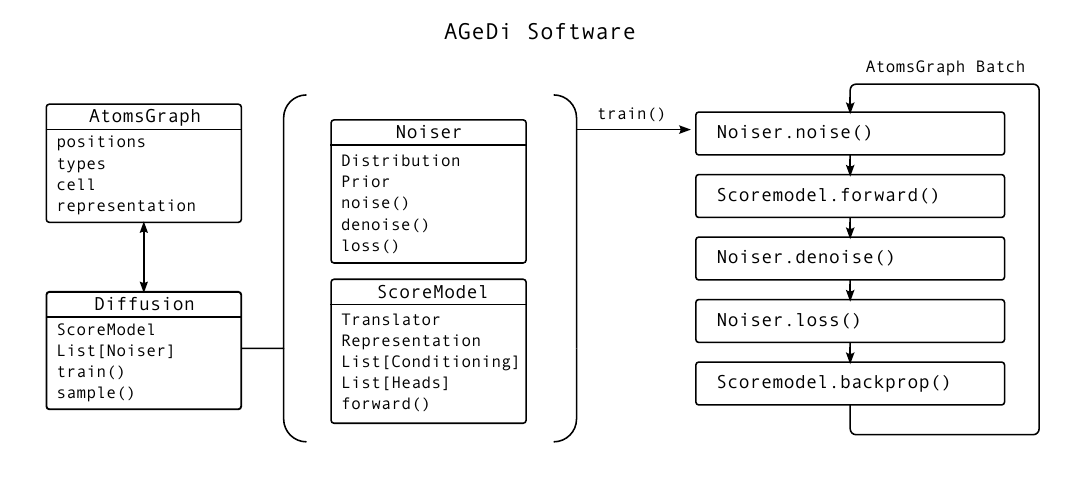}
	\caption{The core classes of AGeDi with the most important
          attributes and methods. Arrows indicate how they interact
          during training of a diffusion model.}
	\label{fig:agedi}
\end{figure*}

AGeDi (pronounced ``a Jedi'') is an open-source Python package that implements continuous-time
generative diffusion models for atomistic systems. It is designed with
a modular, object-oriented architecture, enabling users to compose and
customize models for a wide range of materials modeling tasks. The
software separates concerns across distinct components — graph
representations, score models, diffusion objectives, and training
loops—making it both extensible for research and practical for
integration into simulation workflows. 

Below, we describe the purpose of the important classes shown in
Fig.~\ref{fig:agedi} and how it connects to the theoretical framework
introduced earlier. 

The \texttt{AtomsGraph} class represents the atomistic system,
$\mathcal{M}$. It calculates graph connectivity on demand and can cache it
for efficiency. A key feature of \texttt{AtomsGraph} is its use of a
\texttt{Representation} object, which computes an equivariant representation of
the atomic system. This representation is computed by the \texttt{ScoreModel},
then passed back to \texttt{AtomsGraph} before score prediction — allowing the
model to condition on time and other properties. 

The \texttt{ScoreModel} interfaces with an equivariant GNN capable of
producing symmetry-respecting representations of 
atomic structures. It supports multiple prediction heads, enabling the
model to output scores for different variables, such as atomic
positions or atomic types. The class is also designed to interoperate
with external software through a \texttt{Translator} interface, which can be
implemented for compatibility with specific codebases. Currently,
the PaiNN\cite{schutt2021} equivariant GNN is interfaced with more GNNs planned
to be implemented. The forward call first computes the
atomic representation, then computes and appends all
conditioning-embeddings and finally computes the final scores with
each head.

The \texttt{Noiser} class encapsulates both the forward (noising) and
reverse (denoising) process used in training and sampling. It
maintains modularity by separating the 
definitions of noise distributions and prior
distributions. Additionally, the \texttt{Noiser} implements the appropriate loss
functions for training the score networks.

The \texttt{Diffusion} model integrates functionality from all the above
components and implements the training and sampling routines. It is
designed to work natively with ASE\cite{hjorthlarsen2017}, making it easy to incorporate into
existing materials discovery workflows.

When training diffusion models that operate on multiple atomic
variables (e.g., positions and types), AGeDi uses a weighted sum of
the individual loss terms from each \texttt{Noiser}. This allows users to
control the relative importance of each component—for example,
emphasizing position diffusion over atomic type diffusion when
appropriate.

This modular design makes AGeDi not only a reproducible implementation
of the models described in this work, but also a flexible research
platform for developing new generative approaches for atomistic
systems.

\subsection{Evaluating Generative Models}
\label{sec:eval}

We adopt the precision–recall (PR) framework introduced by Sajjadi et
al.\cite{Sajjadi2018} to evaluate generative diffusion models, as it explicitly
separates two key aspects of generative performance: fidelity and
coverage. Given a reference distribution $P$ (the training data) and a
learned distribution $Q$ (the generated samples), precision quantifies
the fraction of samples from $Q$ that lie near $P$, reflecting sample
quality, while recall measures the fraction of $P$ that is captured by
$Q$, reflecting diversity or mode coverage. The resulting PR curve
traces the Pareto frontier of achievable precision–recall tradeoffs. 

\begin{figure*}
	\centering
	\includegraphics[]{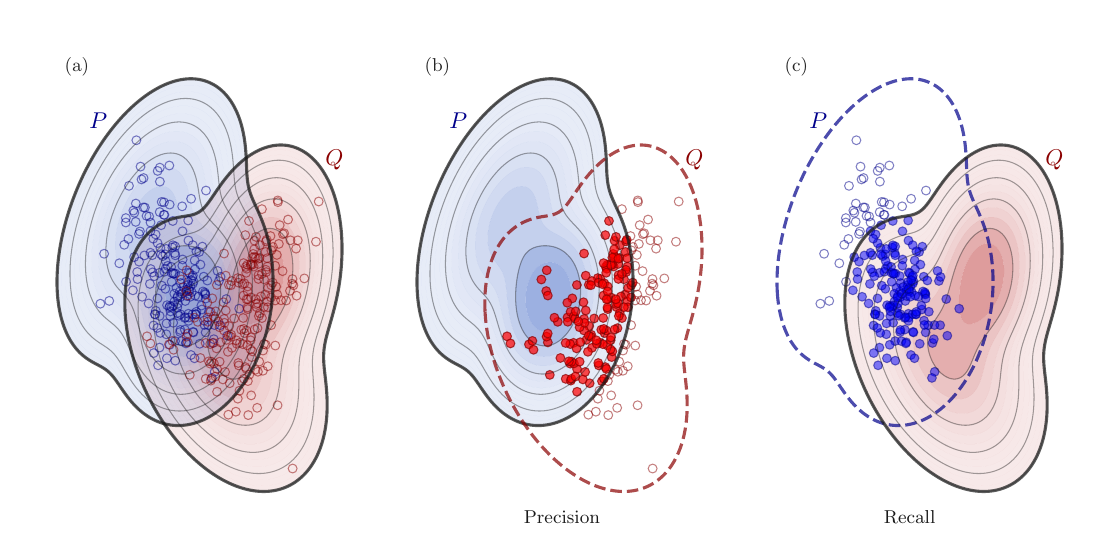}
	\caption{Illustration of the precision and recall metrics. (a) Training
          data from a distribution $P$ and generated samples from
          a learned distribution $Q$. (b) Precision captures sample
          quality i.e. samples being near $P$ distribution. (c)
          Recall captures coverage of training data by a learned
          distribution $Q$. The precision-recall curve traces the
          Pareto frontier of achievable precision-recall tradeoffs.}
	\label{fig:pr}
\end{figure*}

This framework is particularly well-suited to generative modeling of
atomistic systems, as it disentangles failure modes that are often
confounded in scalar metrics. For
instance, mode collapse manifests as high precision but low
recall—indicating that generated samples are individually realistic
but lack structural diversity. Conversely, poor fidelity appears as
high recall with low precision — suggesting the model samples broadly
but fails to capture fine-grained physical accuracy. An intuitive illustration
of precision and recall for generative models is presented in Fig.~\ref{fig:pr}.

To aid interpretation of the precision–recall curves, we
introduce synthetic baselines by applying controlled perturbations to
the training data. These baselines simulate known degradations in
structure fidelity and serve as physical reference points against
which model performance can be compared. Specifically, we randomly
subsample a set of structures from the training set and add Gaussian
noise to their atomic positions. The noise is applied independently to
each atom, with zero mean and varying standard deviations to produce a
range of distortion levels. Importantly, the chemical composition and
overall topology of each structure remain unchanged—only the geometry
is perturbed. 

By generating multiple synthetic datasets with varying noise
levels and varying degrees of subsampling (e.g. $100\%$ and $50\%$ of training
data coverage), we create a family of PR curves that correspond to
progressively degraded versions of the ground truth data. These curves
represent quantifiable benchmarks for interpreting model outputs: for
example, if a generative model achieves precision and recall
comparable to a synthetic set with small added noise, it suggests that
the generated structures closely resemble realistic atomic
configurations. Conversely, if model outputs fall below the baseline
with high noise, it indicates a significant loss in structural
fidelity or diversity. 

This approach provides a physically grounded reference frame for
evaluating generative quality, by situating model performance relative to known
perturbations of real data, we can more meaningfully assess the
trade-off between sample quality and diversity. 

\section{Results}
\label{sec:results}
We demonstrate the capabilities and versatility of AGeDi through two
application-driven case studies, illustrating its performance across
both structural and compositional domains. These experiments validate
the expressiveness of AGeDi’s generative diffusion models, as well as
the modularity of its software stack for addressing distinct modeling
tasks: sampling atomic clusters and sampling two-dimensional materials. We also
highlight the flexibility of the framework by showcasing interpolation
over atomic types to sample out-of-domain bimetallic clusters and
employing CFG to sample two-dimensional materials with
targeted symmetries. 

\subsection{Cluster AGeDi Model}
\begin{figure*}
	\centering
	\includegraphics[]{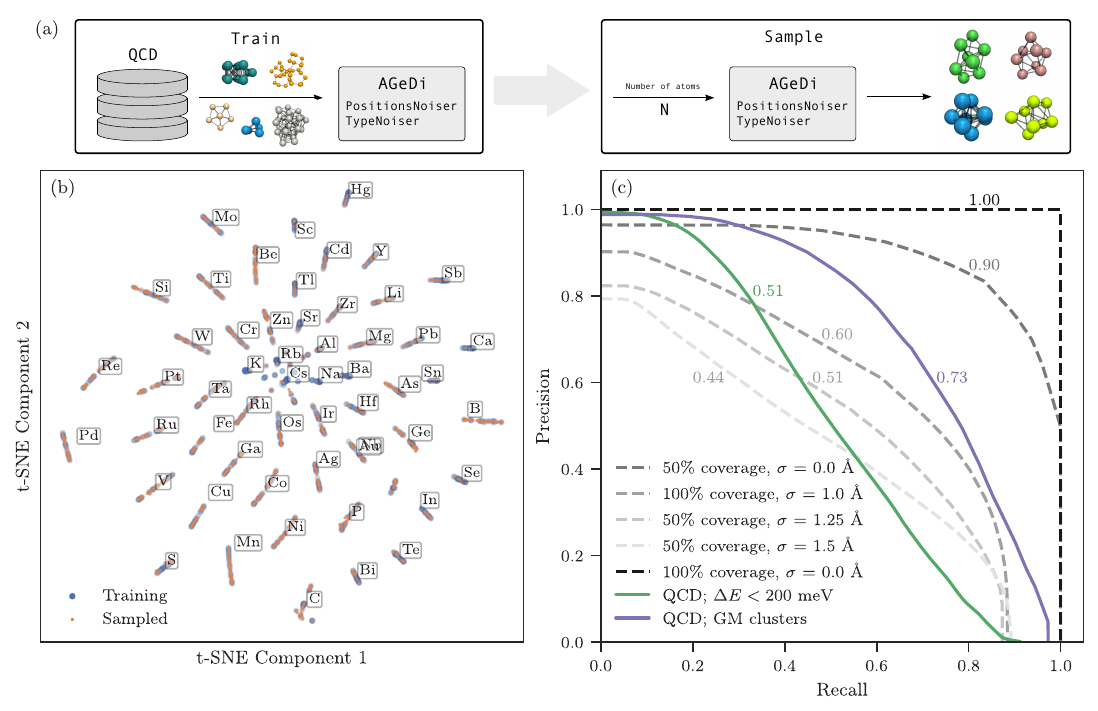}
	\caption{(a) Schematic illustration of the training and
          sampling pipeline. (b) Two-dimensional representation of training structures and
          structures sampled with the trained AGeDi model. (c)
          Precision-Recall curve for the two AGeDi model trained on
          small QCD GM clusters and on
          QCD structures within $200$ meV of the minimum energy
          cluster respectively. The black dashed line represents
          perfect recovery of the training data. The gray dashed lines
          represents synthetically created baselines created by adding Gaussian
          noise to the atomic positions with standard deviation $\sigma$. Each PR-curve is
          labelled with its calculated area under the curve.}
	\label{fig:qcd}
\end{figure*}

Metal clusters are of significant interest in materials science, as
they serve as a bridge between isolated atoms and bulk matter,
exhibiting size-dependent electronic, structural, and chemical
properties. These characteristics offer keys insights 
into nanomaterials, catalysis, and the design of functional
materials.\cite{luo2016}

We use the Quantum Cluster Database, which
only contains low-energy mono-metallic nanoclusters. Two subsets are
constructed for training: 
\begin{enumerate}
\item The global energy minimum (GM) cluster for each stoichiometry with fewer than 30 atoms.
\item All clusters within 200 meV of the ground-state structure, with
  no size restriction.
\end{enumerate}

We train a joint diffusion model on both positions and atomic types,
allowing sampling of clusters by only specifying the number of
atoms. Due to the non-periodic nature of nanoclusters, we employ the
variance-exploding (VE) SDE formulation and standard Gaussian noise. A
schematic illustration of the training and 
sampling pipeline is presented in Fig.~\ref{fig:qcd}(a).

Fig.~\ref{fig:qcd}(b) shows a 2D t-SNE embedding on
SOAP\cite{bartok2013,himanen2020} features of the GM clusters training data and an
equal number of samples from the trained model on the same dataset. The generated samples
appear to broadly cover the same configuration space as the training
data. We also observe that only mono-metallic cluster are sampled,
and thus the type diffusion has correctly learned the training data
atomic type distribution. 
To quantitatively evaluate generative performance, we compute
precision–recall curves for both models (Fig.~\ref{fig:qcd}(c)). We also
generate synthetic baselines by subsampling the training data and
perturbing positions with Gaussian noise of varying standard
deviation, $\sigma$. These
synthetic datasets act as anchors for interpreting the model’s
fidelity and coverage. 
The model trained on the GM clusters dataset shows higher precision and
recall, as expected, given that each stoichiometry appears only once
in this set. The larger $\Delta E < 200$ meV dataset contains more structural
diversity and larger clusters, making it more difficult to model,
which is reflected in lower recall values. 

\begin{figure*}
	\centering
	\includegraphics[]{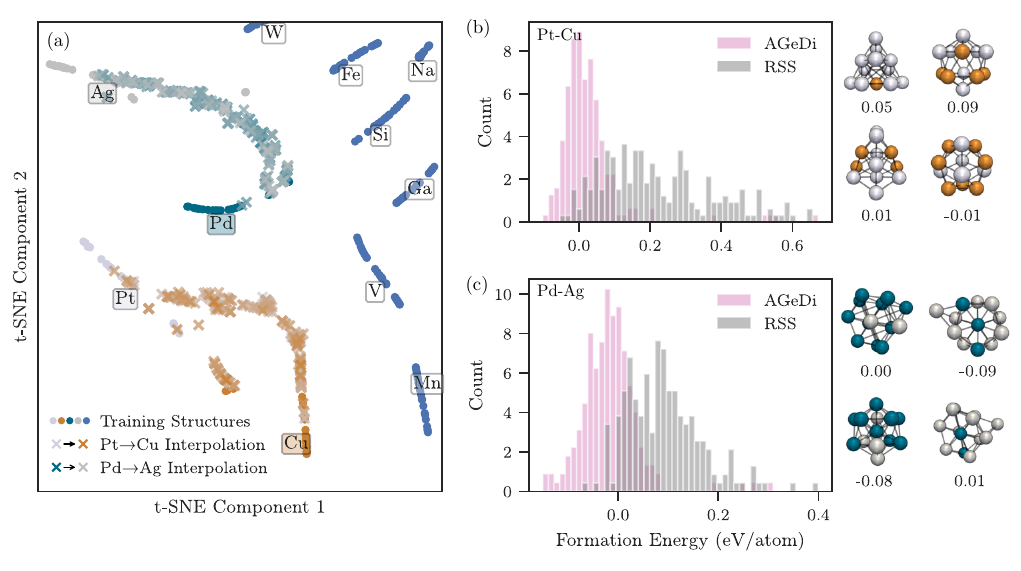}
	\caption{(a) Two-dimensional representation of training structures
          and interpolated Pt-Cu and Pd-Ag clusters. The colors
          interpolate between those of the pure metals and indicate
          the relative proportion of each metal in the sampled
          cluster. (b) Formation energy distribution of sampled Pt-Cu
          clusters compared to RSS. (c) Formation energy distribution of sampled Pd-Ag
          clusters compared to RSS. For each case, four randomly
          selected clusters sampled using the AGeDi model are shown,
          with their formation energies listed beneath each
          structure.}
	\label{fig:interpolation}
\end{figure*}

Even though training is performed on mono-metallic clusters, we
can demonstrate that interpolating atomic type embeddings enables the
generation of plausible bimetallic clusters using the same model. We
follow the type interpolation scheme introduced in section
\ref{sec:interpolation}. Specifically, we target:

\begin{enumerate}
\item Pt$_{12-n}$Cu$_{n}$ ($0 < n \leq 12 $)
\item Pd$_{13-n}$Ag$_n$ ($0<n\leq 13 $),
\end{enumerate}
which are systems that have been widely studied in the literature.\cite{mejia-lopez2009,zhao2013,chaves2015,wei2022,wei2022a}
Fig.~\ref{fig:interpolation}(a) shows a 2D t-SNE embedding of the sampled bimetallic clusters
alongside the training data. The colors
interpolate between those of the pure metals and indicate the relative
proportion of each metal in the sampled cluster. The interpolated
clusters span the configuration space between their mono-metallic
endpoints.

In order to quantify the quality of the interpolated
clusters, we evaluate their potential energy using density functional
theory (DFT) and compare to sampling using a standard
structure search method; namely Random Structure Search
(RSS)\cite{pickard2011,christiansen2022}. To compare energetics across 
stoichiometries, we evaluate the formation energy as
\begin{equation}
  E_{\text{formation}} = \frac{E_{\text{A}_n\text{B}_m} - n
  E_{\text{A}}^{\text{ref.}} -  m E_{\text{B}}^{\text{ref.}}}{n+m},
\end{equation}
where $E_{\text{A}_n\text{B}_m}$ is the sampled cluster and
$E_{\text{A/B}}^{\text{ref.}}$ is the reference energy per atom chosen
as the monometallic cluster of the same size taken from the QCD.
All clusters are fully relaxed with the pretrained MACE-MPA-0
potential\cite{batatia2024}, followed by a single-point DFT
calculation used to evaluate the formation 
energy. For details on DFT see the Supplementary Information. 

Fig.~\ref{fig:interpolation}(b,c) shows the distribution of formation
energies for the Pt-Cu and Pd-Ag clusters respectively along with
examples of the AGeDi sampled clusters. We observe significantly lower
formation energies for the AGeDi sampled clusters compared to
RSS. This result highlights the strong generalization ability of
the trained AGeDi model. In particular, the effective interpolation scheme
enables the model, trained exclusively on monometallic clusters, to
successfully extend its learned structural knowledge to the
generation of stable bimetallic clusters.

\subsection{2D Material AGeDi Model}

\begin{figure*}
	\centering
	\includegraphics[]{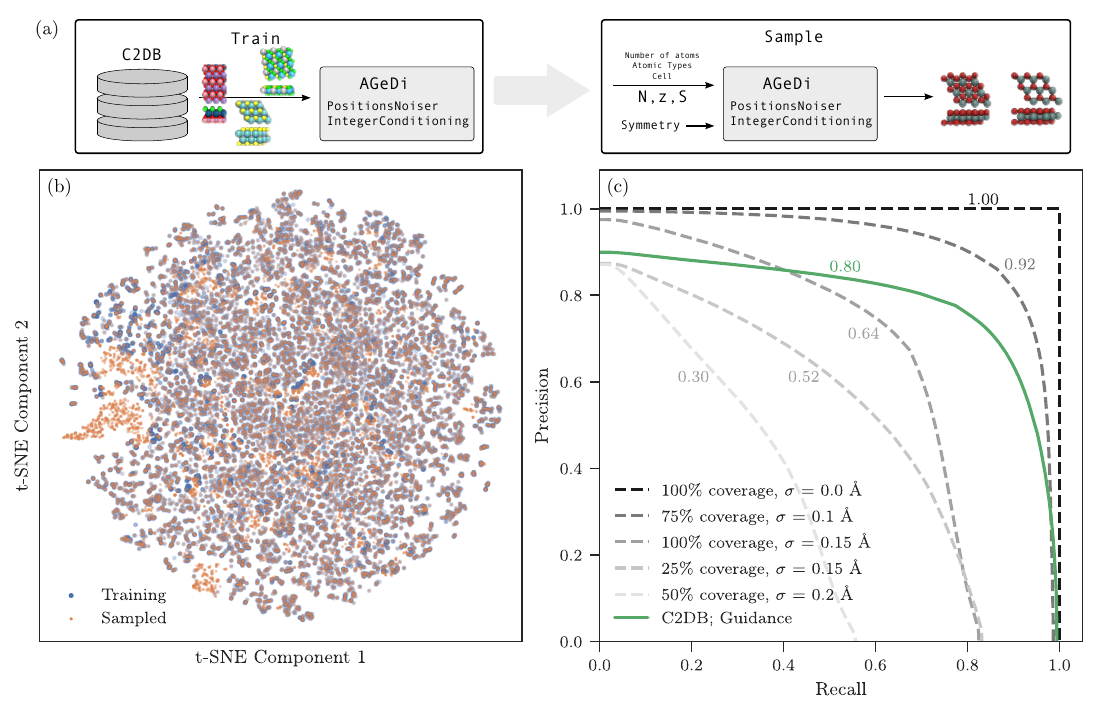}
	\caption{(a) Schematic illustration of the training and
          sampling pipeline. (b) Two-dimensional representation of training structures and
          structures sampled with the trained AGeDi model. (c)
          Precision-Recall curve for AGeDi model trained on
          C2DB database. The black dashed line represents
          perfect recovery of the training data. The gray dashed lines
          represents synthetically created samples for comparison.}
	\label{fig:c2db}
\end{figure*}

Two-dimensional materials have attracted considerable attention
due to their exceptional physical, electronic, and mechanical
properties, which arise from their reduced dimensionality. As
atomically thin systems, they offer a platform for exploring novel
quantum phenomena and hold promise for applications in electronics,
optoelectronics, and energy storage.\cite{lin2023}

The Computational 2D Materials Database
is a large two-dimensional materials database coverage over 30 different
crystal symmetries including their structural, thermodynamic and
magnetic properties. We randomly split the dataset into training and
test subsets.

We train a diffusion model over atomic positions, conditioned on
symmetry, using classifier-free guidance. Following Fu et al.\cite{fu2024}, we use layer group
numbers (1–80) for symmetry labels, which are embedded into a
64-dimensional vector appended to the atomic representation before
score prediction. A schematic illustration of the training and
sampling pipeline is presented in Fig.~\ref{fig:c2db}(a).

Fig.~\ref{fig:c2db}(b) presents a 2D t-SNE projection on MACE-MPA-0
features\cite{batatia2024} comparing training data and
model samples, showing good overlap in structural space. Fig.~\ref{fig:c2db}(c)
shows the PR curve for this model, alongside several synthetic
baselines generated by subsampling and noising the training data.
We observe strong precision and recall, with lower noise levels than
in the cluster case, indicating more consistent and stable structural
patterns across the C2DB dataset. 

\begin{figure*}
	\centering
	\includegraphics[]{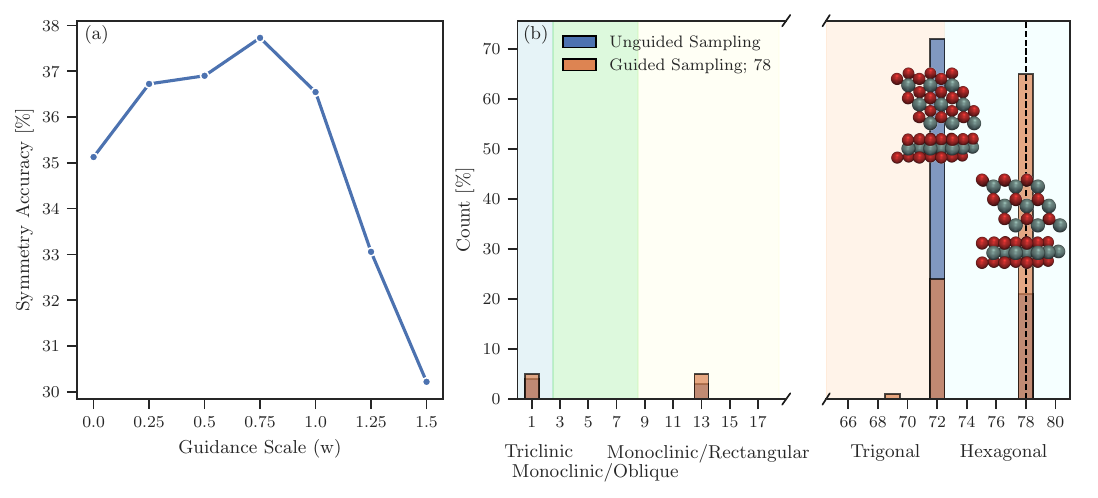}
	\caption{(a) Symmetry accuracy for samples of the C2DB test
          dataset for varying guidance scale, $w$. (b) Histograms of
          SnBr2-materials resulting from unguided and p$\bar{6}$m2 symmetry
          group guided sampling.
        }
	\label{fig:guidance}
\end{figure*}

To assess the effect of classifier-free guidance, we sample structures
from the test set while varying the guidance scale $w$. Fig.~\ref{fig:guidance}(a)
shows the layer group symmetry accuracy i.e. how closely the layer
group symmetry of the sampled structures match the test set. Interestingly, we find that moderate guidance scales ($w \approx 0.5–0.75$)
yield the highest symmetry fidelity in the generated two-dimensional
materials. Contrary to the typical assumption that stronger guidance
improves conditional accuracy, increasing the guidance scale beyond 
$w=1$ leads to a degradation in symmetry preservation. This suggests
that excessive guidance may overconstrain the generation process,
forcing the model into regions that nominally
align with the conditioning signal (e.g., target symmetry class) but
violate physical or geometric consistency. Since symmetry in atomistic
systems emerges from precise spatial relationships rather than being
explicitly labeled, we posit that low to moderate guidance allows the
model to respect both the learned physical priors and the conditioning
constraint, whereas high guidance may distort atomic arrangements in a
way that breaks true symmetry. These findings underscore the
importance of carefully tuning the guidance strength in generative
diffusion models for physics-constrained domains.

Fig.~\ref{fig:guidance}(b) provides an illustrative example: the
SnBr$_2$ system, which can naturally adopt either p$\bar{3}$m1 or
p$\bar{6}$m2 symmetry. Without guidance, 
the model predominantly samples the lower symmetry group. With
targeted guidance, however, it successfully generates p$\bar{6}$m2
structures—demonstrating the model’s ability to steer generation
toward underrepresented symmetries.

\section{Discussion}
\label{sec:discussion}
AGeDi is designed to serve as a flexible and extensible platform for
the community to explore generative modeling in atomistic materials
science. Its integration with ASE and modular architecture make it
readily usable within existing simulation workflows, supporting tasks
such as structure generation, dataset augmentation, and rapid
prototyping of new generative models. By supporting both continuous
and discrete diffusion process, AGeDi accommodates a wide range of
system types — from clusters to surface-supported thin-films
— and can be adapted to specialized domains. The framework also facilitates reproducibility and
benchmarking. The standardized evaluation
metrics (PR curves) and model interfaces make it
a solid foundation for comparative studies and methodological
development in atomistic generative modeling. 

Looking forward, AGeDi will be extended to support diffusion over
the periodic cell, enabling its application to bulk crystalline materials
and variable-cell systems. This enhancements aim to broaden
AGeDi’s utility across a growing range of materials discovery
applications including crystal structure prediction.

\section{Conclusion}
\label{sec:conclusion}
We have introduced AGeDi, a flexible and extensible software package
for training and deploying generative diffusion models tailored to
atomistic systems. The framework supports both continuous and discrete
diffusion process, enabling simultaneous modeling of atomic
positions and chemical species. Notably, AGeDi incorporates
continuous-time discrete diffusion for atomic types, classifier-free
guidance for conditional sampling, and interpolation over atomic type
embeddings — features that extend its
utility beyond standard generative models. 

To assess generative quality, we propose a precision–recall evaluation
framework with synthetically constructed datasets, providing a
more nuanced view of sample fidelity and coverage than 
scalar metrics. 

We validate AGeDi by training two foundational diffusion models: one
for metallic clusters using the QCD dataset and another for two-dimensional
materials using C2DB. These models demonstrate AGeDi's ability to
generate realistic and diverse atomic configurations, perform
interpolation across chemical compositions, and steer generation
toward specific structural symmetries. 

Together, these capabilities establish AGeDi as a practical and
forward-looking tool for data-driven materials discovery, inverse
design, and generative modeling of complex atomistic systems.

\section{Code availability}
The AGeDi package is publicly available
\url{https://github.com/nronne/agedi} under a GNU GPLv3
license. Documentation is available at
\url{https://agedi.readthedocs.io}. 

\section{Acknowledgements}
We acknowledge support from VILLUM FONDEN through Investigator grant,
project no. 16562, and by the Danish National Research Foundation
through the Center of Excellence “InterCat” (Grant agreement no:
DNRF150).

\bibliography{bib}

\section{Supplementary Information}

\subsection{AGeDi Models}
Training and sampling scripts are available at
\url{https://agedi.readthedocs.io} along with a description of
hyperparameters for both QCD and C2DB models. Checkpoints of the
trained models are also available. 

\subsection{Two-Dimensional Representation of Structures}
The two-dimensional representation for the cluster structures are
produced using t-SNE dimensionality reduction on a SOAP
representation.\cite{bartok2013,himanen2020} The two-dimensional
representation for the two-dimensional materials are produced using
t-SNE dimensionality reduction on the MACE-MPA-0
descriptors.\cite{batatia2024}

\subsection{DFT}
DFT is performed with the GPAW\cite{mortensen2024} software using the
Perdew-Burke-Ernzerhof\cite{perdew1996} exchange-correlation functional and a
$400$ eV plane-wave cutoff. 

\section{RSS}
We perform RSS with the AGOX\cite{christiansen2022} software following
their documentation using the MACE-MPA-0\cite{batatia2024} pretrained potential.

\subsection{Precision-Recall Curve Evaluation}
The precision-recall curve are evaluated using the code at\\
\url{https://github.com/msmsajjadi/precision-recall-distributions}.

\subsection{Evaluation of 2D Material Symmetries}
The layer group symmetry is calculated using the spglib software with
\texttt{symproc=0.1}.\cite{togo2024a}

\end{document}